\documentclass[10pt, conference]{IEEEtran}
\IEEEoverridecommandlockouts
\usepackage{amsmath,amssymb,amsfonts}
\usepackage{graphicx}
\usepackage{xcolor}

\begin{document}

\def\baselinestretch{0.95}

\title{Efficient Implementation of RISC-V Vector Permutation Instructions}
\author{
\IEEEauthorblockN{Vasileios Titopoulos, George Alexakis}
\IEEEauthorblockA{\footnotesize Electrical and Computer Engineering\\ 
Democritus University of Thrace, Greece}
\and
\IEEEauthorblockN{Chrysostomos Nicopoulos}
\IEEEauthorblockA{\footnotesize Electrical and Computer Engineering\\  University of Cyprus, Cyprus}
\and
\IEEEauthorblockN{Giorgos Dimitrakopoulos}
\IEEEauthorblockA{\footnotesize Electrical and Computer Engineering\\ 
Democritus University of Thrace, Greece}}

\maketitle

\begin{abstract}
RISC-V CPUs leverage the RVV (RISC-V Vector) extension to accelerate data-parallel workloads. In addition to arithmetic operations, RVV includes powerful permutation instructions that enable flexible element rearrangement within vector registers --critical for optimizing performance in tasks such as matrix operations and cryptographic computations. However, the diverse control mechanisms of these instructions complicate their execution within a unified datapath while maintaining the fixed-latency requirement of cryptographic accelerators. To address this, we propose a unified microarchitecture capable of executing all RVV permutation instructions efficiently, regardless of their control information structure. This approach minimizes area and hardware costs while ensuring single-cycle execution for short vector machines (up to 256 bits) and enabling efficient pipelining for longer vectors. The proposed design is integrated into an open-source RISC-V vector processor and implemented at 7 nm using the OpenRoad physical synthesis flow. Experimental results validate the efficiency of our unified vector permutation unit, demonstrating that it only incurs 1.5\% area overhead to the total vector processor. Furthermore, this area overhead decreases to near-0\% as the minimum supported element width for vector permutations increases.
\end{abstract}


\section{Introduction}
The increasing deployment of Artificial Intelligence (AI) and Machine Learning (ML) applications has triggered the need to accelerate them directly in hardware. AI/ML algorithms often involve performing the same operations on large sets of data (vectors or matrices). Vector processors excel at this by enabling a single instruction to operate on multiple data elements simultaneously, significantly accelerating computations~\cite{khadem2023vector}. 

The RISC-V "V" Vector (RVV) Instruction Set Architecture (ISA) extension includes a variety of data parallel operations that add vector capabilities to the RISC-V CPUs~\cite{rvv}. To enable data rearrangement within vector registers, it also provides a set of powerful permutation instructions.
These instructions facilitate arbitrary merging, grouping, and other complex data shuffling or data copying. This capability is crucial for optimizing performance in diverse kernels~\cite{vector-for-sort}, such as matrix multiplication (for sparse or dense data), matrix transpositions~\cite{avx-perm}, and cryptographic primitives~\cite{rvv-for-crypto}. These in-vector register permutations complement indexed and strided loads, which can rearrange data within vector registers during memory fetching~\cite{padua}.

Each class of vector permutation instructions uses different control information to specify how vector data elements should be rearranged in the destination register. For example, {\tt vcompress} uses a binary mask to select which elements of the source vector register should be gathered at one end of the destination register while preserving their original relative order. Here, the control information describing the permutation directly relates to the input data elements. Conversely, in {\tt vrgather}, the control information refers to the destination register elements; each destination element specifies which input element should be placed in its position.

The differentiation in semantics of control information for vector permutation instructions, both per-input and per-output, complicates their execution within a single datapath. This challenge is further intensified by the requirement for a fixed and data-independent execution latency, to protect against side-channel attacks in crypto accelerators. This work addresses these challenges by designing a unified hardware architecture capable of executing all RVV vector permutation instructions with the same latency, regardless of their control information structure, thereby reducing area and hardware cost.

The focus is on the design of an efficient RVV-compliant hardware vector permutation unit for short-vector length machines (with vector lengths up to 256 bits) that operate on a unified multi-ported vector register file driving 
Single Instruction Multiple Data (SIMD) capable execution units. 
This design aims for single-cycle or pipelined execution of vector permutation instructions with a fixed latency of one or two cycles. 
To accommodate operations on longer vectors using register grouping, 
vector permutations can be executed as a series of single-vector permutation operations~\cite{padua}. The main contributions of this work are summarized as follows:
\begin{itemize}
\item A unified hardware architecture is presented that handles the execution of all vector permutation instructions, despite the difference in the semantics of the control information of each class of vector permutations. Vector permutations follow the selected element width and support 8- and 16- and 32-bit words.
\item The proposed permutation unit ensures a fixed execution latency for all types of permutations, thereby meeting the data-independent execution latency requirement. This consistency helps protect against side-channel timing attacks by preventing data-operand-dependent execution latency variations.
\item The proposed permutation unit was integrated into a RISC-V vector processor and implemented as open source using the OpenRoad Physical design flow driven by the ASAP 7 nm technology library. The physical synthesis results show the scalability of the proposed approach and how the permutation hardware cost can be reduced by increasing the minimum element data width supported for permutations. 
\end{itemize}
\section{Permutation in RISC-V Vector Processors}

A vector permutation involves the rearrangement of the elements within a vector register. RISC-V RVV permutation instructions are split in three categories.

\subsection{RVV permutation instructions}

\begin{figure}[t!]
\centering
\includegraphics[width=0.9\columnwidth]{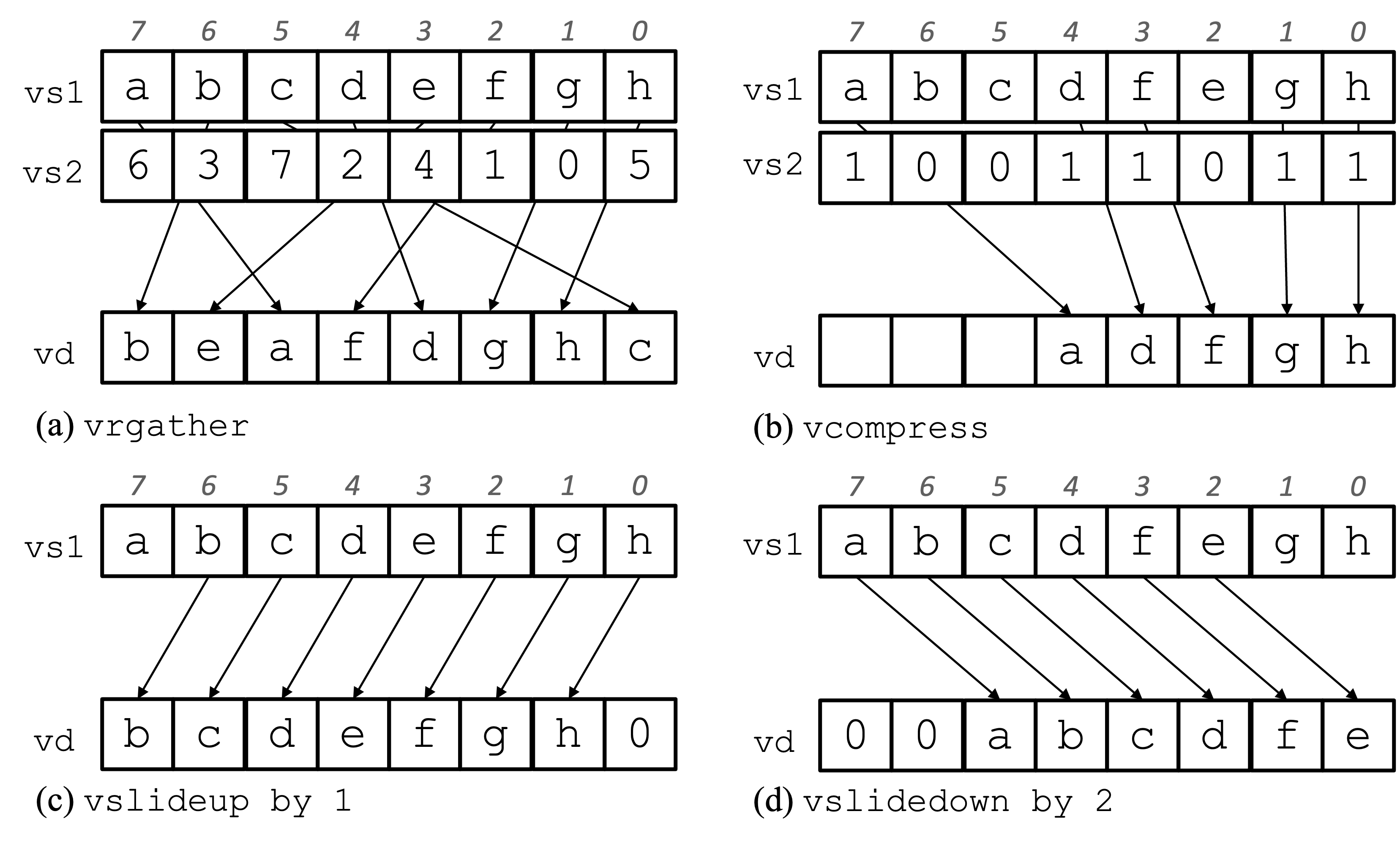}
\caption{An example demonstrating the implementation of (a) {\tt vrgather} and (b) {\tt vcompress}, assuming a vector length of 8.}
\label{f:vrgather_vcompress}
\end{figure}

The {\tt vrgather} instruction offers the most versatile form of vector-element rearrangement.
In this operation, data movement is explicitly determined by per-output indices. The {\tt vrgather} instruction takes two operands: the first is the input vector,  and the second is a vector of indices that dictate 
which input element is selected for each element of the output vector.
An example of the execution of {\tt vrgather} is shown in
Fig.~\ref{f:vrgather_vcompress}(a).

The {\tt vcompress} operation, on the other hand, gathers elements of the input vector into the starting positions of the output vector, based on a mask. The mask identifies which elements in the input vector should be included on the right side of the output vector, without losing their original relative position. An example of the execution of {\tt vcompress} is shown in Fig.~\ref{f:vrgather_vcompress}(b).

The {\tt vslide} instructions are also input-driven permutations and provide two main variants: {\tt vslidedown} and {\tt vslideup}. These operations shift the elements of the input vector in the corresponding direction ({\tt vslidedown} shifts to the right and {\tt vslideup} shifts to the left), based on an offset value. Examples of the execuction of {\tt vslideup} and {\tt vslidedown} are shown in Figs~\ref{f:vrgather_vcompress}(c) and ~\ref{f:vrgather_vcompress}(d), respectively.

All-three permutation instructions can be masked using register {\tt v0}, thus allowing the targeted data re-arrangement to affect only a subset of the output elements.

\subsection{State-of-the-art implementations}
\label{ss:related}

The implementation of vector permutation instructions has taken many forms in recent RISC-V vector processors.

Ara~\cite{cavalcante2019ara} supports the {\tt vslideup} and {\tt vslidedown} instructions, while more complex permutation operations, such as vector gathering and compressing, are implicitly implemented through indexed loads.
This approach minimizes the cost of vector permutation hardware units, but incurs latency proportional to the vector register size. Other processors adopting this strategy include ZeroVex~\cite{zhao2024zerovex}, Vecim \cite{wang202430}, LEM~\cite{fang2022lem}, and newer versions of Ara~\cite{perotti2022new, perotti2024ara2}.

Processors like Virtuvius~\cite{minervini2023vitruvius+}, big.VLITTLE~\cite{ta2022big}, and Spatz~\cite{cavalcante2023spatz} employ an inter-lane interconnect to enable communication-intensive operations (e.g., permutations, reductions). Ring networks are often preferred for their simplicity, allowing lanes to exchange data via the network. 
AraXL~\cite{purayil2025araxlphysicallyscalableultrawide} also uses a ring network to connect to neighbor vector clusters to perform large vector permutations.

To execute a {\tt vrgather} instruction in a network requires the transformation of the per-output source indices of the {\tt vs2} register into per-input destination indices that can be handled by the network's routing.  
For this reason, two ring networks were employed in~\cite{valente2020vector}: a data network and an index network. The index network converts source indices to destinations by routing element index {\tt i} to position {\tt vs2[i]}, enabling subsequent data network transfers. It should be noted that the use of networks for permutations makes permutation execution latency variable and data dependent.

Contrary to previous approaches, Vicuna~\cite{platzer2021vicuna}, Saturn~\cite{zhao2024instruction}, and Ocelot -- a vector engine for the BOOM processor~\cite{zhaosonicboom} --  
permute one element per cycle (similar to loading an element per cycle in an indexed load).
LEM~\cite{fang2022lem} follows a similar strategy, but it also uses additional index registers to manage memory accesses and permutations.

\section{The micro-architecture of the unified datapath}
\label{s:architecture}
The design of a unified datapath for vector permutation instructions is complicated by the fundamental difference in control semantics between \textit{input-driven} and \textit{output-driven} approaches.

Since the RISC-V RVV ISA extension allows for re-configurable vector element width, the presented permutation unit is designed to support the smallest possible element width (1 byte), and wider widths are treated as series of consecutive bytes. In this way, all possible vector element widths are fully and seamlessly supported.

\subsection{The execution of the {\tt vrgather} instruction}
\begin{figure}[t]
\centering
\includegraphics[width=0.9\columnwidth]{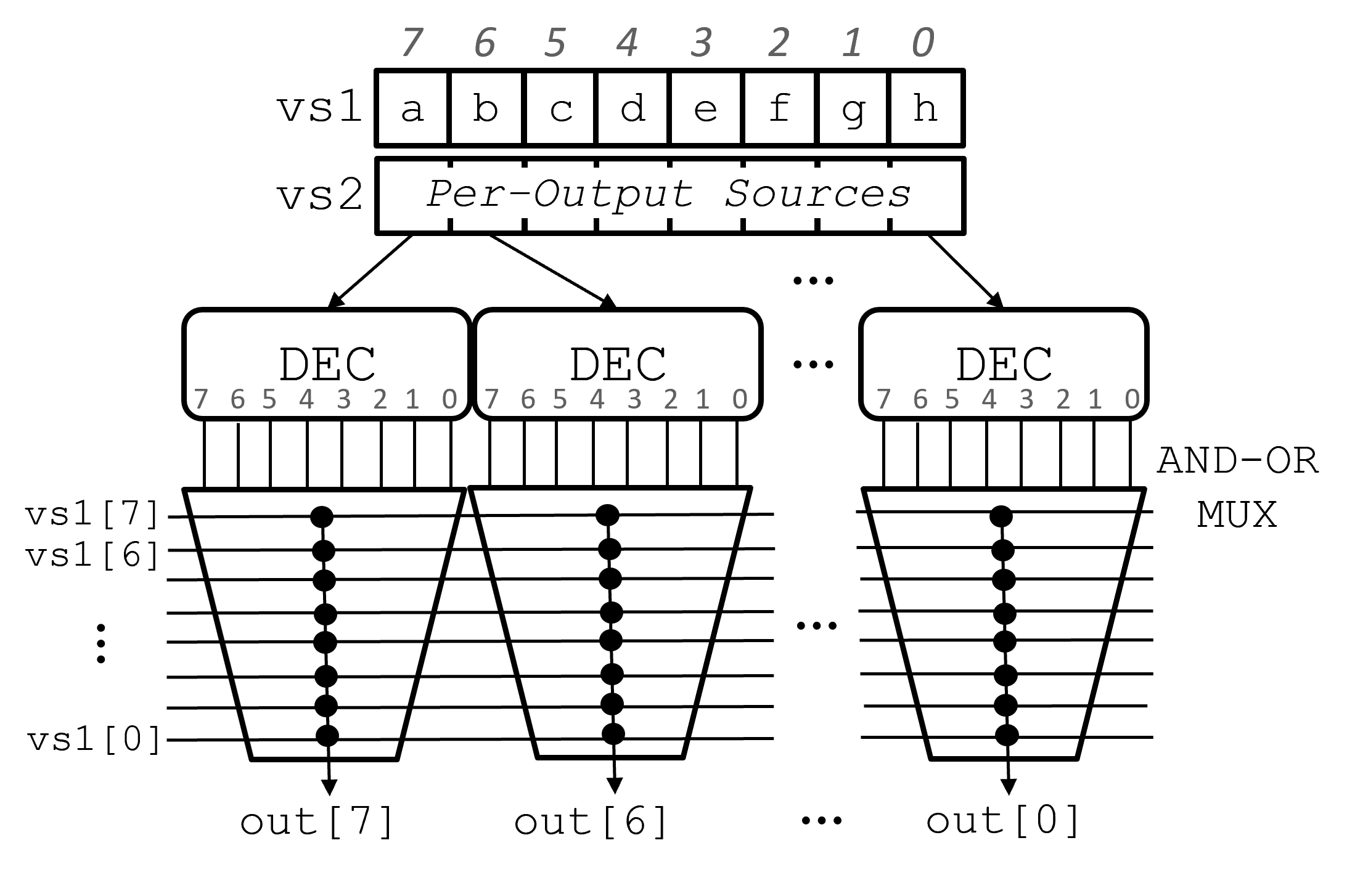}
\caption{The crossbar logic consisting of an AND-OR multiplexer per output element that executes {\tt vrgather} for an 8-element vector. The per-output source addresses are first decoded (to one-hot) before driving the output multiplexers.}
\label{f:vrgather-xbar}
\end{figure}

The proposed architecture for a unified vector permutation unit relies on a crossbar with per-output AND-OR multiplexers, as shown in Fig.~\ref{f:vrgather-xbar}. This structure is innately amenable to output-driven permutation instructions, such as {\tt vrgather}. The select signals of each output of the crossbar are simply the one-hot equivalent vectors of the value in the corresponding element in the vs2 register of the instruction. 
Note that, as per the definition of {\tt vrgather}, this setup allows for the same input element to be copied to multiple output elements.

This datapath can handle all \textit{output-driven} permutation instructions. The next step is to devise a mechanism to facilitate the execution of all \textit{input-driven} permutations -- e.g., {\tt vcompress}, {\tt vslide}, etc. -- on the same datapath with minimal overhead/complexity.

\subsection{Executing vector compress on the same crossbar logic}
 The control bits in the second source operand ({\tt vs2}) of input-driven instructions include mask bits that refer to the corresponding elements in the first source operand ({\tt vs1}), and these bits cannot be directly related to a specific output element. Thus, those bits must be converted into suitable information that can leverage the existing \textit{per-output} crossbar structure. 
To achieve this, we follow a two-step approach. First, the mask bits in {\tt vs2} are transformed into a per-input, output-element destination. This means that, for each input element of {\tt vs1}, we determine the corresponding output element to which it should move. Second, to \emph{reuse the crossbar that executes} {\tt vrgather}, we must rearrange the destination indices into a format compatible with the crossbar logic.

\subsubsection{From mask bits to per-input output destinations}
\begin{figure}[t]
\centering
\includegraphics[width=\columnwidth]{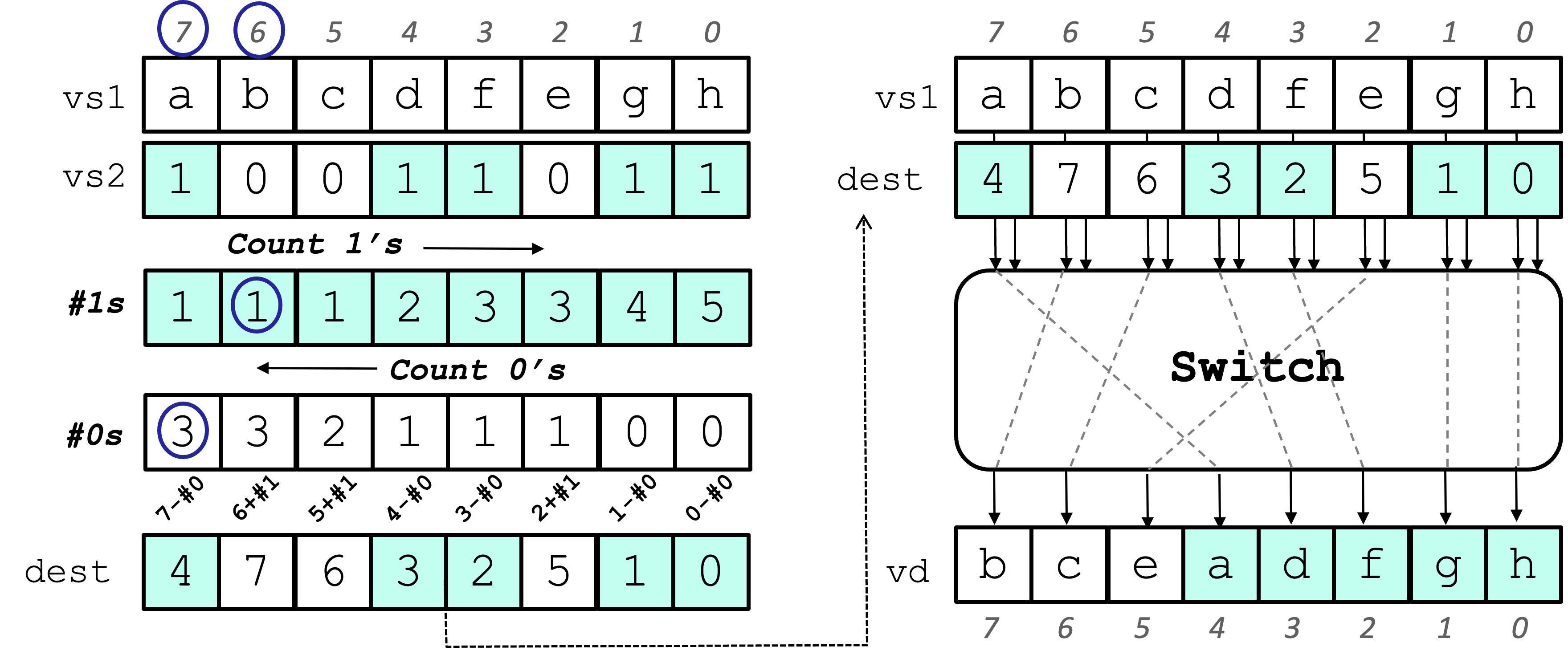}
\caption{The transformation of the mask control bits of {\tt vcompress} results in per-input destination indices that indicate which output element the corresponding input should move to.}
\label{f:vcompress-dest}
\end{figure}

The transformation algorithm comprises two operations: the calculation of two vectors of prefix sums and the addition/subtraction of these sums to/from the corresponding vector element indexes (positions). The whole transformation flow is illustrated in Fig.~\ref{f:vcompress-dest}. The first vector of prefix sums is generated from left to right (high to low vector indexes) and counts the number of 1s in the bit mask contained in the second operand ({\tt vs2}) of the instruction. The second vector of prefix sums is generated in the opposite direction, i.e., from low to high vector indexes, and counts the number of 0s in vector {\tt vs2}. The last step of the algorithm generates the desired vector of per-input destinations. The value of each element of this vector is calculated as follows: if the corresponding mask bit in {\tt vs2} is 1, then the corresponding prefix sum-of-0s is \emph{subtracted} to its vector index. On the other hand, if the corresponding mask bit in {\tt vs2} is 0, then the corresponding prefix sum-of-1s is \emph{added} from its vector index.

Hence, the final result of this pre-processing algorithm is a vector indicating the destination element (in the output vector register) of each input element in {\tt vs1}. The right part of Fig.~\ref{f:vcompress-dest} shows how this vector of output destinations can be used to implement the {\tt vcompress} instruction. 
Although it may seem unnecessary to move the mask-0 elements to the left, this is done to ensure that no two elements in the vector of destinations have the same value. The importance of this attribute will be explained shortly.

\begin{figure}[th]
\centering
\includegraphics[width=0.5\columnwidth]{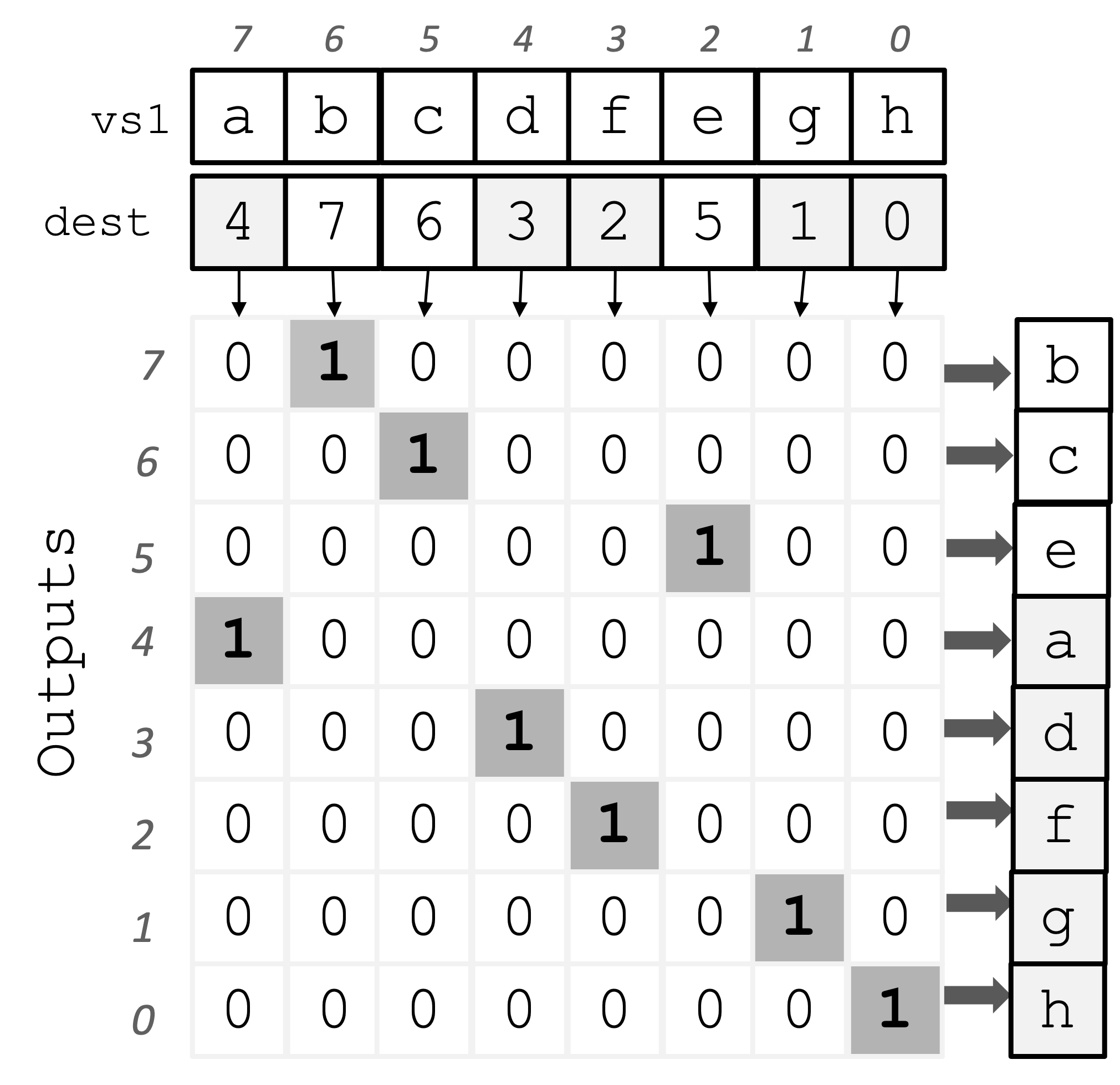}
\caption{The decoded (one-hot) form of the per-input destinations can drive the outputs of a crossbar to move input elemenets according to the functionality of {\tt vcompress}.}
\label{f:vcompress-onehot}
\end{figure}

\subsubsection{From per-input output destinations to one-hot signals for the crossbar}
The elements of this vector can now be used directly as select signals for the same crossbar used by output-driven permutation instructions. Fig.~\ref{f:vcompress-onehot} visualizes how this can be achieved by using the matrix abstraction for the crossbar. As can be seen, each element in the newly generated vector of destinations is converted into a one-hot \textit{vertical} 
vector. 

Same as with output-driven instructions, each \textit{row} still corresponds to the select signals of the corresponding output of the crossbar. Each row of this matrix is guaranteed to also be one-hot (a pre-requisite for functional correctness) by the definition of the input-driven instructions themselves: no two (or more) input elements may go to the same destination element. By construction, the vector of destinations generated by the pre-processing algorithm includes different destination values in each of its elements, as previously mentioned.

\begin{figure}[t]
\centering
\includegraphics[width=0.8\columnwidth]{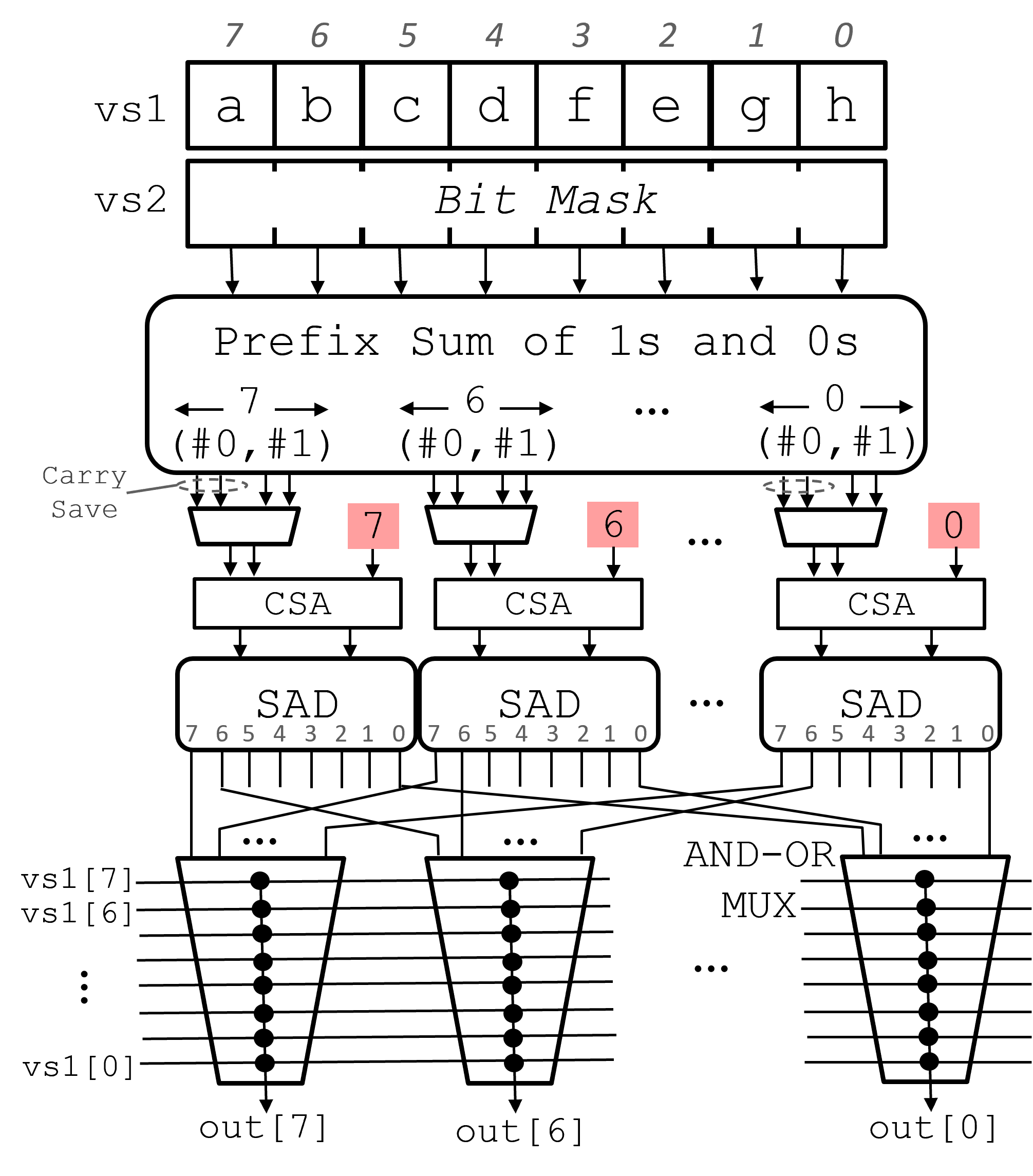}
\caption{The crossbar-based logic that executes the {\tt vcompress} instructions. Sum-Addressed Decoders (SAD) implement fuse addition and decoding without the need for carry propagation.}
\label{f:vcompress-complete}
\end{figure}

\subsubsection{Hardware implementation}
\begin{figure}[ht]
\centering
\includegraphics[width=0.45\columnwidth]{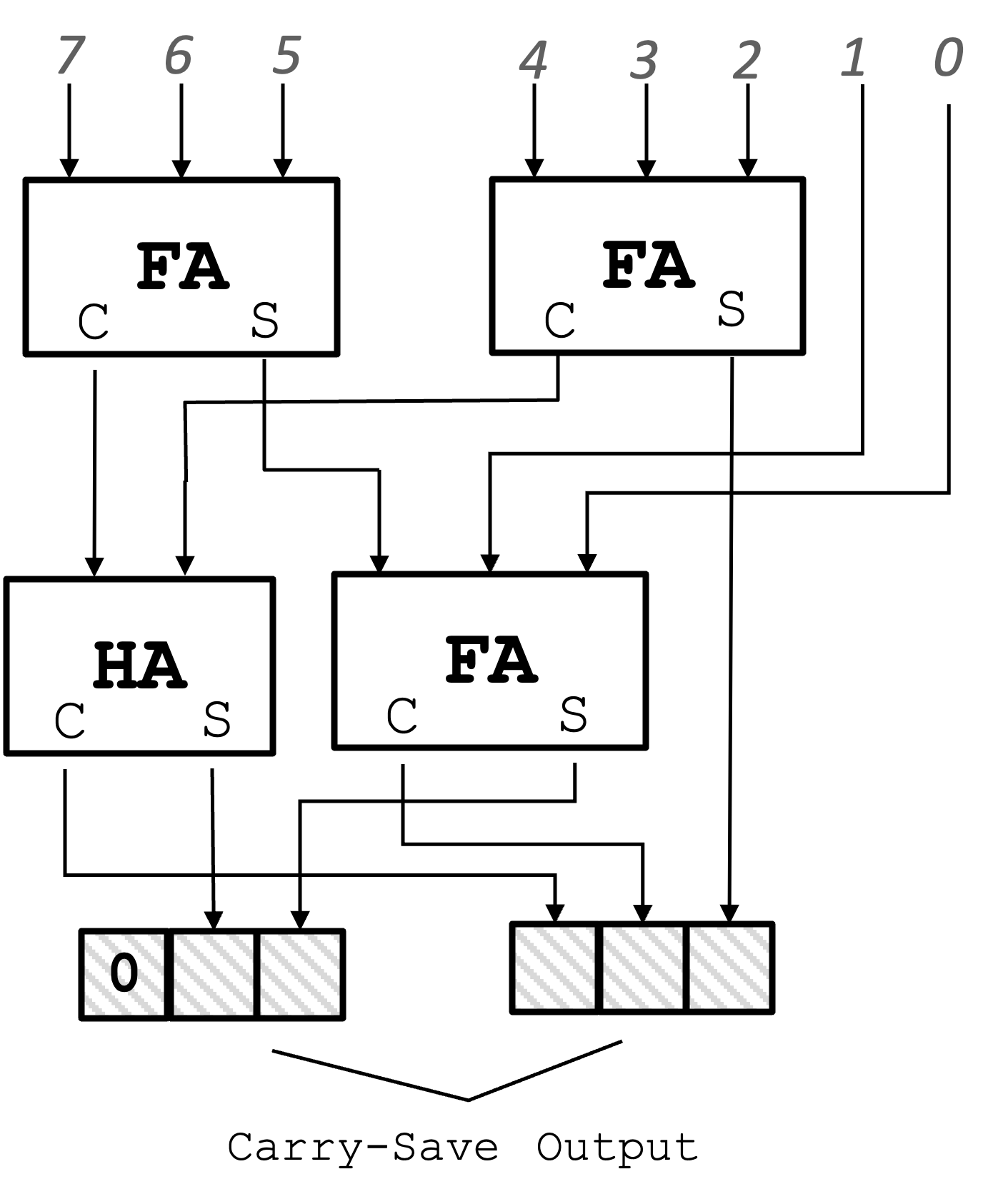}
\caption{Counting the number of 1's from position 7 down to 0 and returning the sum in carry-save format. The redundant representation of the output eliminates the need for internal carry propagation.}
\label{f:counter}
\end{figure}

The hardware implementation of the pre-processing algorithm described above is shown in Fig.~\ref{f:vcompress-complete} for an 8-element datapath. The two prefix sums are computed in parallel using carry-save counters~\cite{counter} at each element position. To eliminate unnecessary delays caused by carry propagation within the counting logic, the prefix sums are maintained in carry-save form. Fig.~\ref{f:counter} illustrates the carry-save counting logic used to count the number of ones from position 7 down to position 0 while keeping the result in carry-save form. Similar cells are used for the remaining positions and also for counting the number of zeros.

Returning to Fig.~\ref{f:vcompress-complete}, the prefix sum at each position must be added to -- or subtracted from -- the index of the corresponding position, and the resulting value should be directly decoded. Since the prefix sum at each position is in carry-save form, it is added to the position index using a Carry-Save Adder (CSA). Since position indexes are fixed values, the CSA logic is simplified after logic synthesis.

The output of the CSA specifies the destination of the corresponding element in a redundant carry-save format. To determine the destination in a non-redundant format, a carry-propagate adder is typically used to sum two outputs of the CSA. Instead of following this approach, the CSA result is passed directly to a Sum-Addressed Decoder (SAD)~\cite{sam,sad}. The SAD computes the final sum while simultaneously converting it into a one-hot vector representation. This add-and-decode operation inherently eliminates the need for internal carry propagation~\cite{sad}. As a result, the final decoded destinations are produced with minimal delay, as carry propagation is avoided at every step.

Each input element requires one such SAD unit to determine its corresponding one-hot destination vector. These vectors correspond to the one-hot vertical vectors shown in Fig.~\ref{f:vcompress-onehot}. After appropriate wire reshuffling of the SADs' output, we derive the horizontal one-hot vectors of Fig.~\ref{f:vcompress-onehot}, which serve as select signals for the crossbar’s output AND-OR multiplexers.

\subsection{Executing vector slides}

\begin{figure}[t]
\centering
\includegraphics[width=0.5\columnwidth]{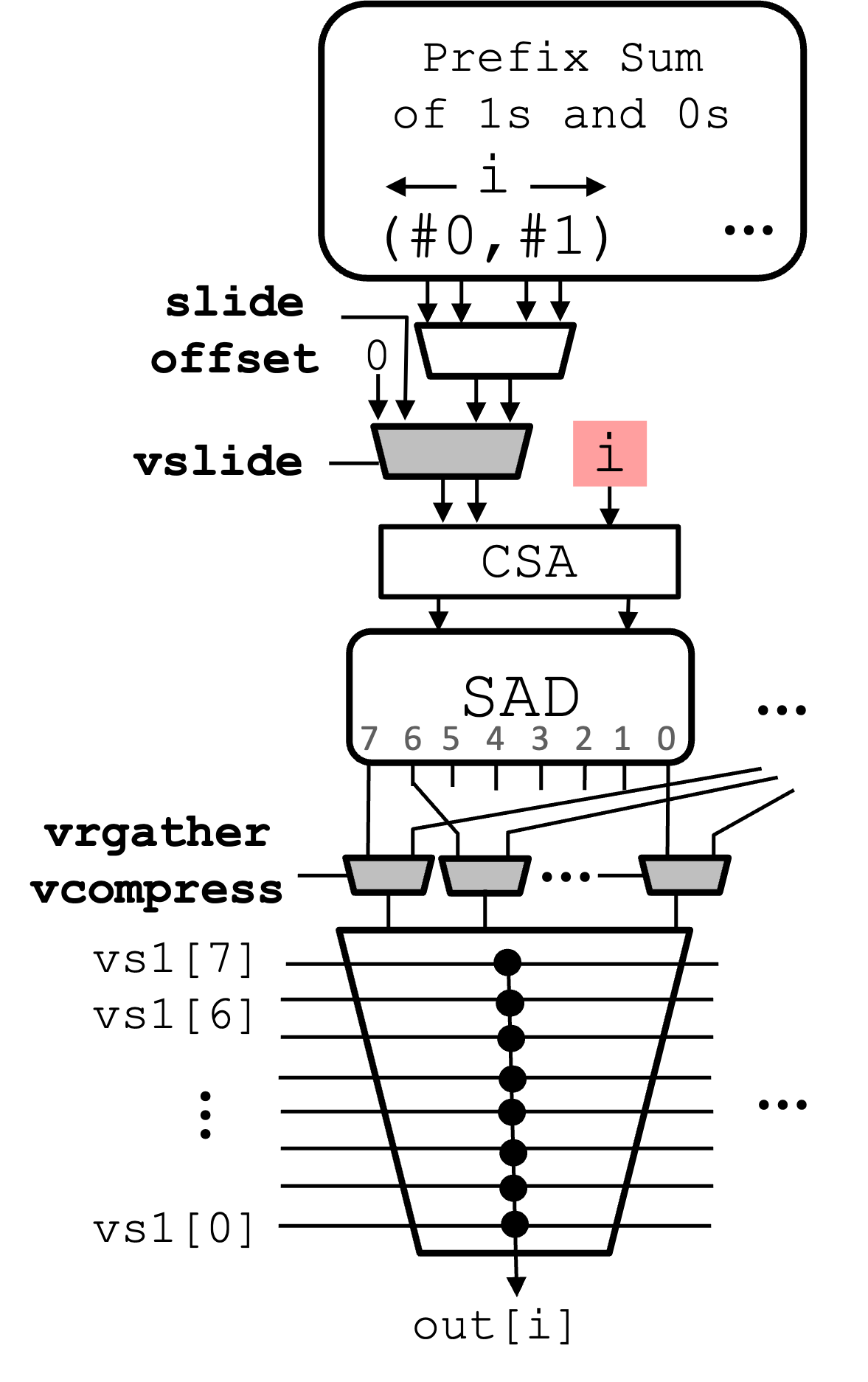}
\caption{The bit slice of the permutation logic at the $i$\textsuperscript{th} output. A positive or negative offset is added to the index of the $i$\textsuperscript{th} bit position and directly decoded via the SAD. In the case of {\tt vrgather}, the decoded signals directly drive the multiplexer of the corresponding output, while in the cases of {\tt vcompress} and {\tt vslide}, the decoded signals are reshuffled to different outputs.}
\label{f:bit-slice-zoom}
\end{figure}

Vector slides can share parts of the datapath presented for the {\tt vcompress} instructions. Specifically, the slide operations of {\tt vslideup} and {\tt vslidedown} do not need the calculation of prefix sums. Instead, the slide offset (included within the instruction itself) is simply added (or subtracted in two's complement) to/from each input element index to calculate the destination in the output register. Hence, the indexes and slide amount are fed directly into the SAD, thereby bypassing the prefix sum logic. 
By construction, if a sum exceeds the vector-element width, then the SAD outputs an all-zeros vector, which correctly ensures that the affected input element is not sent to any output. Such out-of-bounds sums correspond to the input elements that ``slide out'' of the output vector register after the slide operation. The inclusion of the positive or negative slide offset using an additional multiplexer at the input of each SAD is highlighted in Fig.~\ref{f:bit-slice-zoom}. 

Furthermore, since we reuse the same AND-OR multiplexers for {\tt vrgather}, {\tt vcompress}, and {\tt vslide}, the \textit{per-output} and \textit{per-input} select-signal distribution interconnects are multiplexed before reaching the output multiplexers, depending on the class of the permutation instruction (output- or input-driven, respectively). These extra multiplexers are highlighted in gray at the lower part of Fig.~\ref{f:bit-slice-zoom}. 

\section{Evaluation}
\label{s:eval}

\begin{figure}
\centering
\includegraphics[width=0.5\columnwidth]{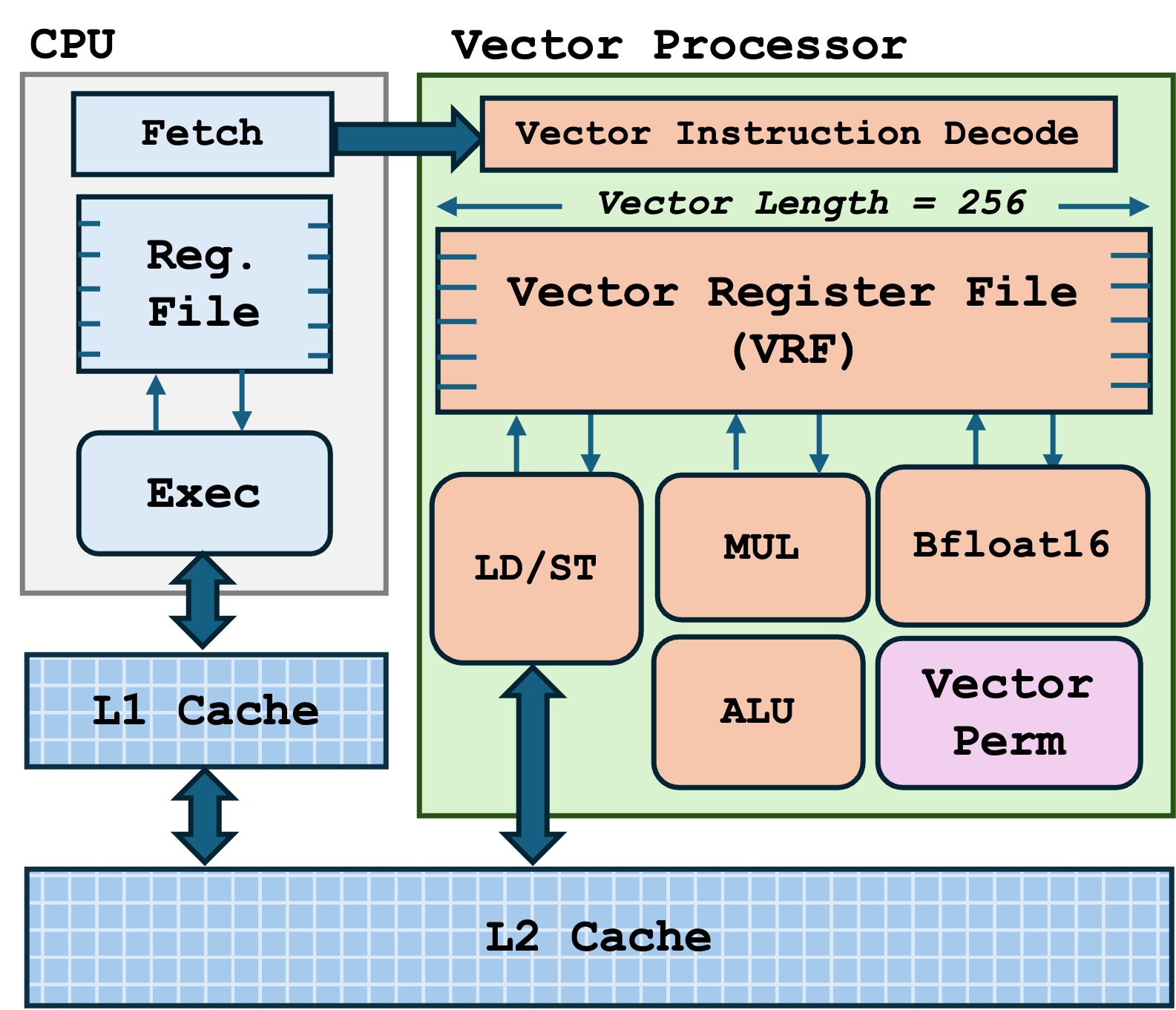}
\caption{The high-level organization of the implemented CPU and the attached vector processor.}
\label{f:cpu}
\end{figure}

The goal of this section is to evaluate the complexity of the hardware implementation of the proposed unified permutation unit. To ensure a realistic evaluation, the proposed permutation unit was integrated into a full-fledged decoupled vector processor, which is attached to a 2-way superscalar out-of-order core. The general organization of the employed processor is shown in Fig.~\ref{f:cpu}. The SystemVerilog RTL of the designed processor is publicly available\footnote{https://github.com/ic-lab-duth/RISC-V-Vector-Processor}.

The vector processor adopts a monolithic organization similar to Saturn~\cite{zhao2024instruction}, featuring a data path that supports a vector length of 256 bits and a unified, multi-ported register file driving a unified cluster of SIMD functional units. In contrast, longer vector architectures employ lane-distributed vector register files and execution units~\cite{perotti2024ara2, minervini2023vitruvius+}. The vector processor includes a vector ALU and a MUL unit for vector integer arithmetic, a vector floating-point unit that operates on Bfloat16 datatypes~\cite{bfloat16}, and a vector permutation unit.

For the comparisons, two versions of the vector permutation unit were designed, both supporting permutations at the byte level, meaning the smallest movable element is one byte. In the baseline version, the vector processor executes: (a) {\tt vrgather} instructions using the crossbar logic shown in Fig.~\ref{f:vrgather-xbar}; (b) {\tt vslide} instructions using a separate slider, which is a logarithmic shifter operating at the byte level; and (c) {\tt vcompress} instructions using a sequential datapath that moves one element -- with asserted mask bit -- per cycle to its correct position in the output vector register (similar to~\cite{zhao2024instruction}). Consequently, {\tt vcompress} instructions are executed as multi-cycle operations. In the second version, the processor includes the proposed unified vector permutation unit, which executes all vector permutation instructions.

\begin{figure}[t]
\centering
\includegraphics[width=0.85\columnwidth]{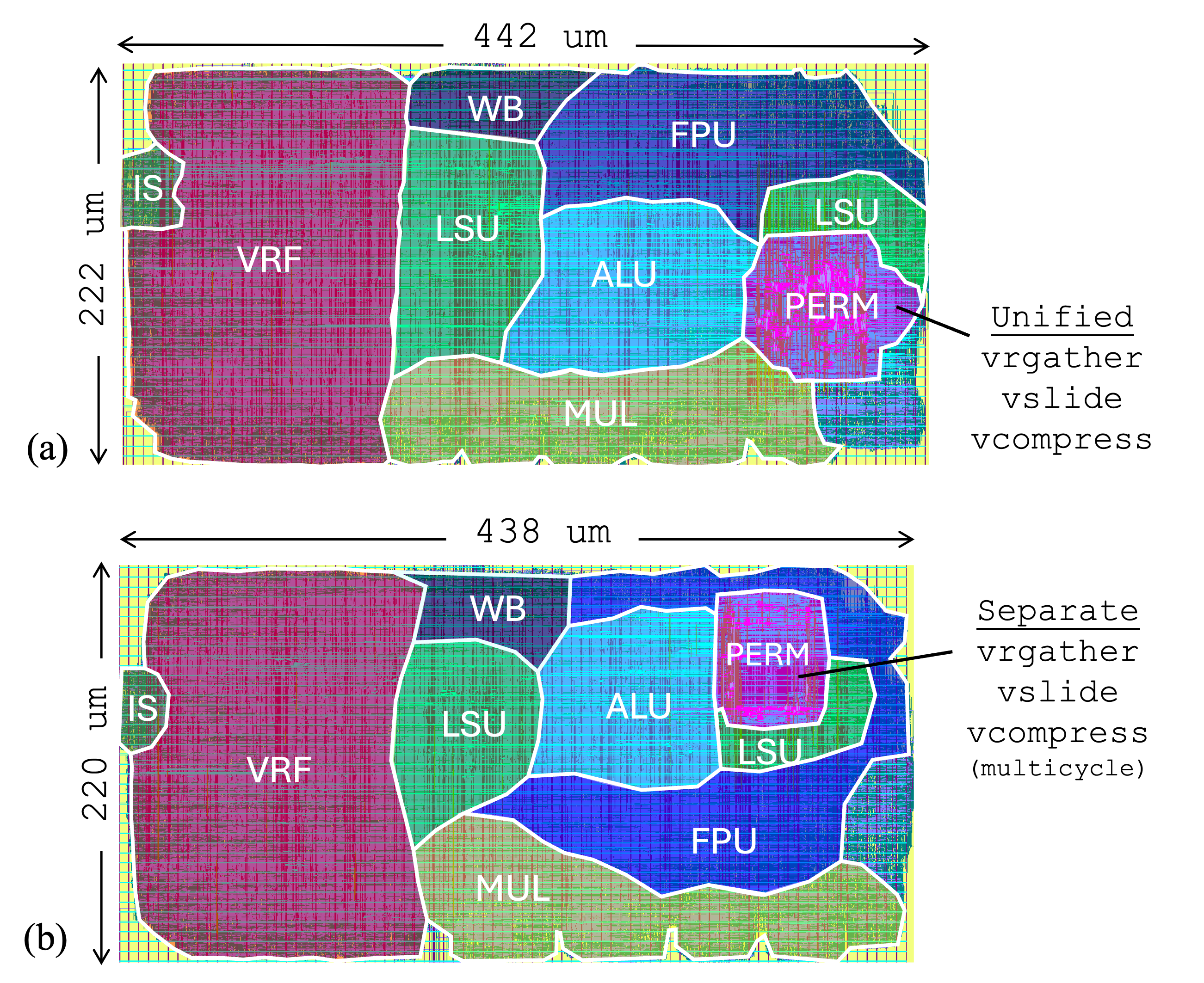}
\caption{The physical layout of both vector processors under comparison: (a) the vector processor that includes the proposed unified permutation unit; and (b) the same vector processor with a permutation unit that includes \textit{separate} datapaths for {\tt vrgather}, {\tt vslide}, and {\tt vcompress}. Compress instructions are executed using multi-cycle operations (one cycle per vector element).}
\label{f:layouts}
\end{figure}

Both designs were synthesized down to the physical layout level using the OpenRoad physical synthesis flow~\cite{openroad}, leveraging the ASAP 7 nm standard-cell library and targeting a clock frequency of 500 MHz under \emph{the same physical design constraints}. The generated physical designs are shown with annotations in Fig.~\ref{f:layouts}. The areas of the two compared vector processors are shown in Table~\ref{t:areas}. 
In both cases, the crossbar-based permutation units fit well within the processor's layout without creating routing congestion issues. The area of the permutation unit is small compared to other vector execution units. For example, the permutation unit corresponds to approximately 40\% of the area of the vector multipliers.

\begin{table}[t]
\centering
\caption{Area of Vector Processor with VL=256 bits at 7 nm}
\label{t:areas}
\begin{tabular}{ccc}
\hline
Size of Element & Baseline & Proposed \\
\hline
1 byte & 96,630 $\text{um}^2$ & 98,124 $\text{um}^2$ \\
2 bytes & 93,537 $\text{um}^2$ & 93,708 $\text{um}^2$  \\
\hline
\end{tabular}
\end{table}

As can be seen in Fig.~\ref{f:layouts}, the proposed permutation unit is around 15\% larger than the baseline permutation unit. However, the permutation unit occupies only a small portion of the vector processor; thus, the incurred area overhead to the \textit{entire vector processor} is about 1.5\% (first row of Table~\ref{t:areas}). This result proves that the proposed address-generation logic added in front of the crossbar and its carry-propagation-free functionality enable high-speed implementations while keeping the area overhead to a minimum. Reaching higher clock frequencies is straightforward by pipelining the indexing and the crossbar logic into different stages, without affecting the constraint of ensuring data-independent execution latency.

To assess the impact of the \textit{size} of the minimum supported \textit{movable element}, the same two designs were re-synthesized with said supported element size increased to two bytes (from one). This ``relaxation'' in the minimum granularity of permutations significantly reduces the area of the permutation units in both cases. The areas of the entire vector processors become almost indistinguishable, as shown in the second row of Table~\ref{t:areas}. 

Power consumption was evaluated using representative vectorized kernels, each containing at least one vector permutation instruction~\cite{rvv-bench}. The switching activity of all nets was quantified through post-layout gate-level logic simulation. Power was then estimated in OpenSTA~\cite{openroad}, by incorporating the extracted wire parasitics from the physical layout. The differences in average power measurements between the two examined vector processors generally align with the differences observed in area. In both cases, the {\tt vrgather} and {\tt vslide} instructions contribute nearly the same amount of power in both the separate and the proposed unified permutation unit. This is because, for both instructions, the carry-save prefix summation of the unified datapath is bypassed. On the other hand, {\tt vcompress} exhibits higher power consumption, which is a direct consequence of its single-cycle implementation, as compared to the sequential nature of its implementation in the baseline processor. For applications that rely only on primitive vector slides of only one position, it is advantageous to execute them outside the main unified datapath. For all other cases, the unified permutation unit is the preferred choice. 

\section{Conclusions}

This work presents an efficient RISC-V RVV-compliant hardware vector permutation unit designed for short-vector-length machines (supporting vectors up to 256 bits). The unit enables single-cycle or pipelined execution of vector permutation instructions with fixed, data-independent latency. The proposed unified datapath executes all vector permutation instructions with a minimum element width of 1 byte, while larger element sizes of 16 and 32 bits are inherently supported. Additionally, the presented carry-propagation-free destination computation logic -- a key component of the new permutation unit -- scales efficiently to larger numbers of inputs. Compared to a vector processor that includes separate datapaths for the {\tt vrgather}, {\tt vslide}, and {\tt vcompress} instructions, this approach executes \textit{all} vector permutation instructions in a single unified datapath with merely a 1.5\% increase in the total vector processor area.

\section*{Acknowledgments}
This work was supported by a research grant from Codasip, a provider of customizable RISC-V IP and Codasip Studio design toolset, and its University Program to Democritus Univ. of Thrace for ``RISCV vector processor design''.

\bibliographystyle{IEEEtran}
\bibliography{references}

\end{document}